\documentclass[a4paper,11pt]{article}
\pdfoutput=1 

\usepackage{jheppub}
\usepackage[T1]{fontenc}

\newcommand{\bea}{\begin{eqnarray}}  
\newcommand{\eea}{\end{eqnarray}}  

\newcommand{\ba}{\begin{array}}
\newcommand{\ea}{\end{array}}
\newcommand{\dis}{\displaystyle}
\newcommand{\lag}{{\cal L}}

\title{
Inverse see--saw neutrino masses in the Littlest Higgs model with T--parity}

\author[]{Francisco del Aguila,}
\author[]{Jos\'e Ignacio Illana,}
\author[]{Jos\'e Mar{\'\i}a P\'erez-Poyatos,}
\author[]{and Jos\'e Santiago}

\affiliation[]{CAFPE and Departamento de F{\'\i}sica Te{\'o}rica y del Cosmos, Universidad
de Granada, E\textendash{}18071 Granada, Spain}

\emailAdd{faguila@ugr.es}
\emailAdd{jillana@ugr.es}
\emailAdd{jmppoyatos@ugr.es}
\emailAdd{jsantiago@ugr.es}

\abstract{We show that the inverse see-saw is the most natural way of implementing neutrino masses in the Littlest Higgs model with T--parity. 
	The three extra quasi--Dirac neutrinos are needed to cancel the quadratically divergent contributions of the 
	mirror leptons to the Higgs mass. 
	If the T--parity of the heavy neutrino singlets is chosen to be even, 
	their contributions to lepton flavor violating transitions are one--loop finite. 
	The most stringent limits on this scenario result from the non--observation of these transitions. 
	Constraints on neutrino mixing imply an upper bound on the mass of the 
	T--odd mirror leptons at the reach of the LHC and/or future colliders.}

\begin{document}
	\maketitle
	\flushbottom
		
	\section{Introduction}
	\label{Introduction}
	
	After the discovery of the Higgs boson \cite{Chatrchyan:2012xdj,Aad:2012tfa}, there are 
	increasing indications of a mass gap to the next physics scale, $f$, above the TeV 
	\cite{gap,gap1}. 
	Such a scenario can be naturally implemented in non-minimal supersymmetric 
	models, as well as in composite Higgs models 
	\cite{Kaplan:1983fs,Dugan:1984hq,Agashe:2004rs}, with the scalar observed 
	at the LHC being a pseudo--Nambu--Goldstone (NG) boson in this latter case. 
	Among this second class of models, the Littlest Higgs model with T--parity (LHT) 
	\cite{ArkaniHamed:2002qy,Cheng:2003ju,Cheng:2004yc,Low:2004xc,Cheng:2005as} 
	emerges as a well--motivated and phenomenologically viable simple model 
	\cite{Hubisz:2004ft,Hubisz:2005bd,Chen:2006cs,Blanke:2006sb,Buras:2006wk,Belyaev:2006jh,Blanke:2006eb,Hill:2007nz,Blanke:2007db,Hill:2007zv,Han:2008wb,Goto:2008fj,delAguila:2008zu,Blanke:2009am,delAguila:2010nv,Goto:2010sn,Zhou:2012cja,Yang:2016hrh,delAguila:2017ugt,delAguila:2019htj}. 
	
	As a matter of fact, neutrino physics remains the only signal of new physics beyond the 
	minimal Standard Model (SM) \cite{Tanabashi:2018oca}. 
	Then, although the LHT is designed to interpret the Higgs boson as a pseudo--NG 
	boson, it must also account for the observed neutrino masses and mixing. 
	As we shall see, T--parity, which plays an essential phenomelogical role suppressing new indirect 
	effects and reducing direct production limits of new (T--odd) particles 
	for they must be pair--produced, also has a significant impact on the mechanism of 
	neutrino mass generation. 
	
	In this paper we show that the LHT can naturally accommodate the inverse see-saw 
	of type I \cite{Mohapatra:1986aw,Mohapatra:1986bd,Bernabeu:1987gr}, 
	and the observed pattern of neutrino masses and mixing \cite{Esteban:2018azc}, 
	without breaking T--parity. Lepton Number (LN) must be explicitly broken 
	at some stage if the observed neutrinos acquire Majorana masses. 
	The LHT has the matter content to account for see--saw mechanisms of type I and II 
	\cite{Minkowski:1977sc,GellMann:1980vs,Yanagida:1979as,Mohapatra:1979ia,Schechter:1980gr,Magg:1980ut,Cheng:1980qt,Gelmini:1980re,Lazarides:1980nt,Mohapatra:1980yp}. 
	However, the type II see--saw, originally considered in the literature \cite{Han:2005nk}, 
	relies on the spontaneous breaking of T--parity \cite{Blanke:2007db}. 
	Even more, the invoked coupling giving neutrinos a mass explicitly breaks T--parity, 
	what implies that it can not be generated by quantum corrections as we argue below. 
	
	In the following we shall show that the minimal lepton content of the model is fixed if one requires that the 
	Higgs mass does not receive quadratically divergent contributions and that Lepton 
	Flavor Violating (LFV) processes, in particular, Higgs decays into two different 
	opposite--charge leptons, remain one--loop finite. 
	This is assuming the originally proposed mechanism for implementing T--parity in the fermion sector 
	and the associated Yukawa Lagrangian giving masses to the T--odd partners of the SM 
	fermion doublets (mirror fermions) \cite{Low:2004xc,Cheng:2005as}. 
	LFV processes stay one--loop finite when we also assume that the SM 
	right--handed (RH) 
	charged leptons are singlets under the global symmetry and that they obtain masses 
	through the usual (minimal) Yukawa interaction 
	\cite{Hubisz:2004ft,Chen:2006cs,Goto:2010sn}. 
	However, the finiteness of these processes is only guaranteed if we require the heavy neutrino 
	singlets completing the RH multiplets under the unbroken global symmetry to be T--even. 
	In this case they mix with SM neutrinos already at tree level and 
	the corresponding mass matrix is the inverse see--saw one, 
	once small Majorana masses are assumed for their heavy left--handed (LH) singlet counterparts. 
	Hence, all phenomenological implications derived from this mechanism 
	follow, in particular, the constraints on the mixing between SM 
	and heavy leptons obtained from Electro--Weak Precision Data (EWPD) and 
	from the non--observation of LFV processes 
	\cite{Abada:2007ux,Arganda:2014dta,DeRomeri:2016gum,Ballett:2019bgd}, 
	as we shall summarize. 
	The conclusion is that, even though T--parity alleviates the flavor problem, 
	we still have to tune the model to reduce the possible misalignment between 
	the SM and the heavy fermions in the absence of an extra flavor symmetry. 
	Constraints on neutrino mixing result in an upper bound on the mass of 
	T--odd mirror leptons, which are at the reach of the LHC and/or future colliders.  
	
	In next section we introduce the notation and justify why the inverse see--saw 
	is naturally implemented in the LHT. In particular, we emphasize that the see--saw 
	of type II must be expected to be suppressed relative to the see--saw of type I. 
	In Section \ref{Constraints} we review the current constraints on the inverse 
	see--saw and the allowed regions of LHT parameters. The last section is 
	devoted to conclusions and final comments on the implications for LHC searches.
	
	\section{Neutrino masses in the LHT}
	\label{Model}
	
	Let us introduce the LHT to fix our notation and assumptions \cite{delAguila:2008zu,delAguila:2017ugt}. 
	(For excellent reviews see \cite{Schmaltz:2005ky,Perelstein:2005ka,Panico:2015jxa}.) 
	The model realizes non--linearly the global $SU(5)$ symmetry which is broken down to $SO(5)$, 
	giving rise to 14 NG bosons 
	\bea
	\Pi=\left(\ba{ccccc}
	-\dis\frac{\omega^0}{2}-\frac{\eta}{\sqrt{20}} & -\dis\frac{\omega^+}{\sqrt{2}} & -i\dis\frac{\pi^+}{\sqrt{2}} & -i\Phi^{++} & -i\dis\frac{\Phi^+}{\sqrt{2}} \\
	-\dis\frac{\omega^-}{\sqrt{2}} & \dis\frac{\omega^0}{2}-\frac{\eta}{\sqrt{20}} & \dis\frac{v+h+i\pi^0}{2} & -i\dis\frac{\Phi^+}{\sqrt{2}} & \dis\frac{-i\Phi^0+\Phi^P}{\sqrt{2}} \\
	i\dis\frac{\pi^-}{\sqrt{2}} & \dis\frac{v+h-i\pi^0}{2} & \sqrt{\dis\frac{4}{5}}\eta & -i\dis\frac{\pi^+}{\sqrt{2}} &  \dis\frac{v+h+i\pi^0}{2} \\
	i\Phi^{--} & i\dis\frac{\Phi^-}{\sqrt{2}} & i\dis\frac{\pi^-}{\sqrt{2}} & -\dis\frac{\omega^0}{2}-\frac{\eta}{\sqrt{20}} & -\dis\frac{\omega^-}{\sqrt{2}} \\
	i\dis\frac{\Phi^-}{\sqrt{2}} & \dis\frac{i\Phi^0+\Phi^P}{\sqrt{2}} &  \dis\frac{v+h-i\pi^0}{2} & -\dis\frac{\omega^+}{\sqrt{2}} & \dis\frac{\omega^0}{2}-\frac{\eta}{\sqrt{20}}
	\ea\right) \ .
	\label{goldstones}
	\eea
	They act on the fundamental representation of the unbroken subgroup 
	multiplying by $\xi = e^{i\Pi/f}$. 
	The action of T--parity is defined to make T--odd all 
	but the SM scalar doublet $\phi = \left( -i\pi^+ \,\; (v+h+i\pi^0)/\sqrt{2} \right)^T$ of hypercharge 1/2 
	($v\simeq 246$ GeV is the SM vacuum expectation value (vev) and the 
	superscript $T$ means transpose): 
	\bea
	\Pi\stackrel{{\rm T}}{\longleftrightarrow}-\Omega\Pi\Omega\ ,\quad 
	\Omega={\rm diag}(-1,-1,1,-1,-1)\ .  
	\label{Pi}
	\eea
	Four of them, $\omega$ and $\eta$, are eaten by the T--odd replica 
	of the electro--weak gauge bosons whereas the other six, $\Phi$, transform 
	as a complex electro--weak triplet of hypercharge 1. 
	
	In the fermion sector each SM lepton doublet 
	$l_L = \left( \nu_L \; \ell_L \right)^T$ is doubled  introducing two incomplete quintuplets \cite{Cheng:2004yc,Low:2004xc} 
	($\sigma^2$ is the second Pauli matrix): 
	\bea
	\Psi_1=\left(\ba{c} - i \sigma^2 l_{1L} \\ 0 \\ 0 \ea\right),\quad
	\Psi_2=\left(\ba{c} 0 \\ 0 \\ - i \sigma^2 l_{2L} \ea\right) , 
	\label{Psimultiplets}
	\eea
	with $\Psi_2$ transforming with the fundamental $SU(5)$ representation $V$ and $\Psi_1$ with its complex conjugated $V^*$, 
	\bea
	\Psi_1\longrightarrow V^*\Psi_1\ ,\quad
	\Psi_2\longrightarrow V\Psi_2\ .
	\eea
	The indices 1 and 2 must not be confused with the family index, which we 
	will omit if not necessary. 
	The action of T--parity on the LH leptons is then defined to be 
	\bea
	\Psi_1\stackrel{{\rm T}}{\longleftrightarrow}\Omega\Sigma_0\Psi_2\ , \quad {\rm with} \quad 
	\Sigma_0=\left(\ba{ccc} 0 & 0 & {\bf 1}_{2\times2} \\
	0 & 1 & 0 \\ 
	{\bf 1}_{2\times2} & 0 & 0 \ea\right) \ .
	\label{TPsi} 
	\eea
	T--parity is thus implemented in the fermionic sector duplicating the SM doublet 
	$l_L = (l_{1 L}-l_{2 L})/\sqrt{2}$, corresponding to the T--even combination 
	$(\Psi_1+\Omega\Sigma_0\Psi_2)/\sqrt{2}$, with an extra heavy mirror doublet 
	$l_{H L} = \left( \nu_{H L} \; \ell_{H L} \right)^T = (l_{1 L}+l_{2 L})/\sqrt{2}$ obtained from the T--odd orthogonal 
	combination $(\Psi_1-\Omega\Sigma_0\Psi_2)/\sqrt{2}$. 
	This extra doublet per family will get its mass combining with a RH 
	doublet $l_{H R}$ in an $SO(5)$ multiplet $\Psi_R$, transforming with the 
	fundamental $SO(5)$ representation $U$, 
	\bea 
	\Psi_R =\left(\ba{c} \psi'_R \\ \chi_R \\ - i \sigma^2 l_{H R} \ea\right) , \quad 
	\Psi_R&\longrightarrow U\Psi_R\ . 
	\label{complete}
	\eea
	The non--linear Yukawa coupling generating this large mass $\sim f$ reads 
	\bea
	\lag_{Y_H} = -\kappa f \left(\overline\Psi_2\xi+ 
	\overline\Psi_1\Sigma_0\xi^\dagger\right)\Psi_R
	+{\rm h.c.}\ , 
	\label{mirror}
	\eea
	where the first term preserves the global symmetry for $\xi {\rightarrow} V \xi U^\dagger$. 
	While the second one is its T--transformed once the T--transformed of $\Psi_R$ is fixed to be $\Omega\Psi_R$ \cite{Cheng:2005as,Pappadopulo:2010jx,Balkin:2017yns}. 
	
\begin{figure}
		\centering
		\begin{tabular}{ccc}
			\includegraphics[scale=0.35]{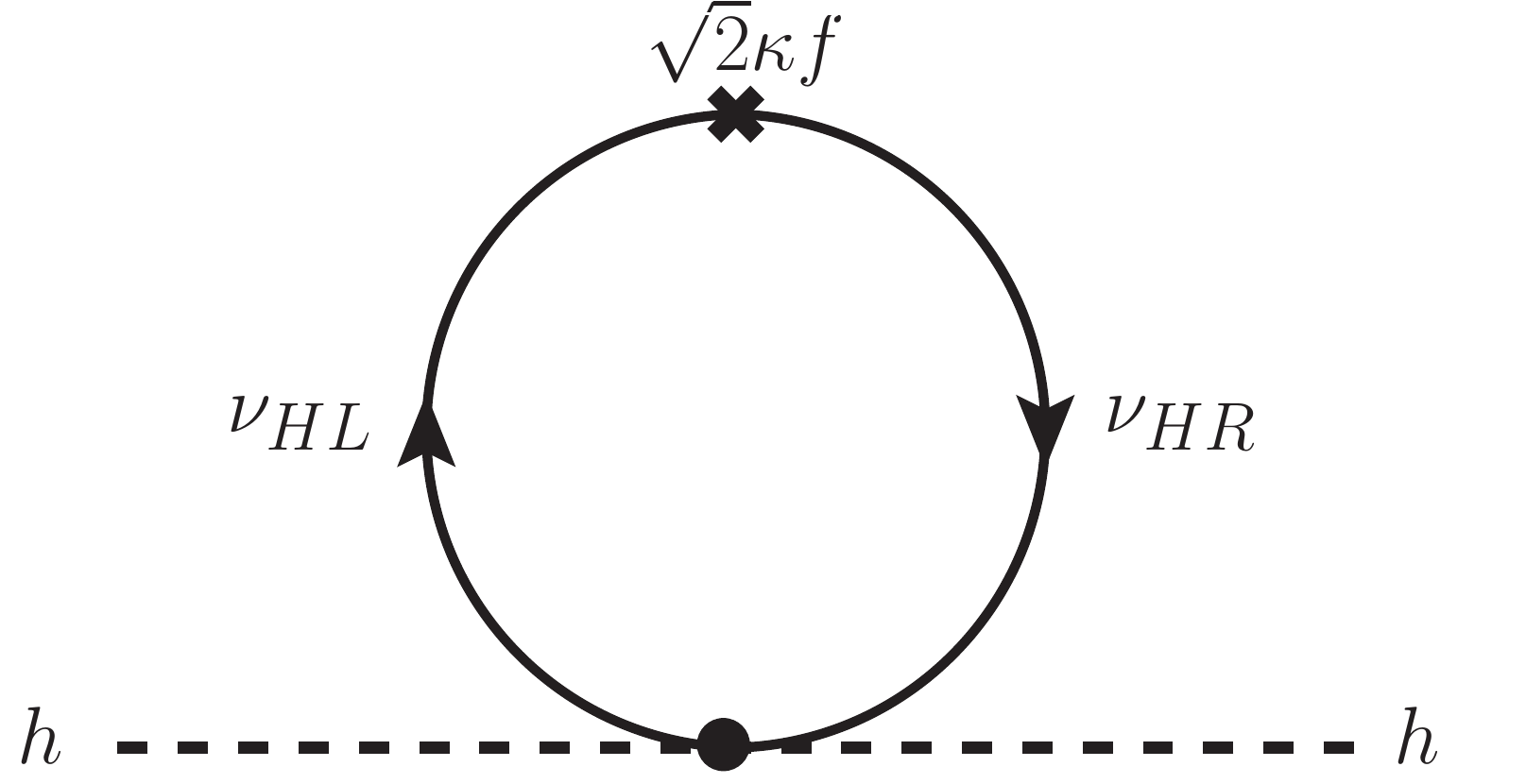} \quad
			& \quad $\begin{array}{c}  \\[-1.5cm] {\bf +} \\[1.5cm] \end{array}$ \quad & \quad
			$\begin{array}{c}  \\[-1.4cm]\includegraphics[scale=0.38]{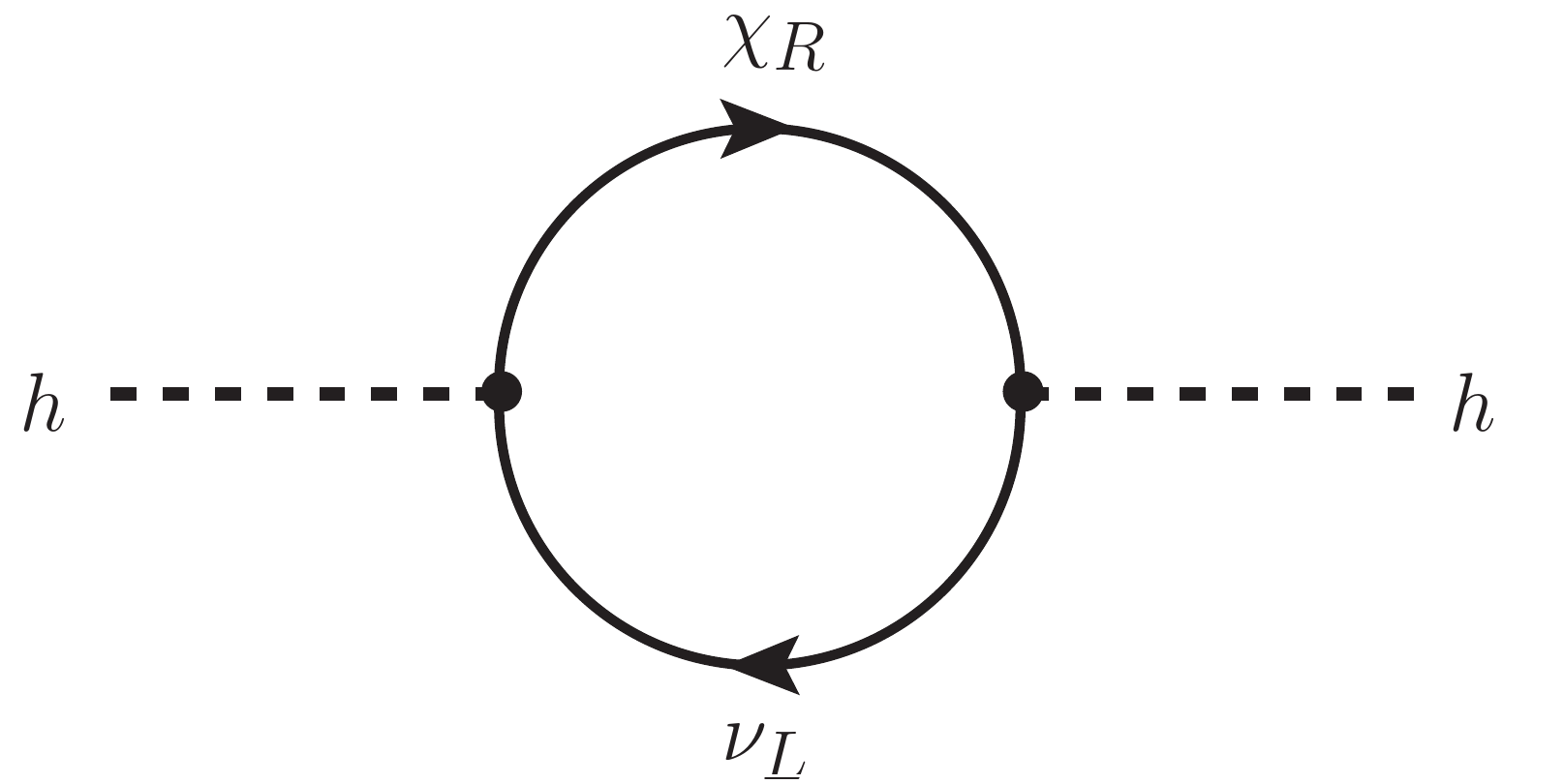}\\[1.4cm] \end{array}$
		\end{tabular}
		\caption{Diagrams contributing to the Higgs mass. The mirror lepton exchange (left) 
			cancels the mirror-singlet mixing contribution (right).}
		\label{diagrams1}
	\end{figure}

	This Yukawa Lagrangian then constrains the heavy fermion content, also restricting the see--saw pattern, 
	as we discuss in the following. 
	Besides giving a vector--like mass $\sqrt{2} \kappa f$ to $\nu_{H}$, it also 
	gives a quadratically divergent contribution to the Higgs mass through the diagram 
	in Fig. \ref{diagrams1} (left). 
	This contribution with a mass insertion and $\nu_{H L, R}$ running in the loop is cancelled by the 
	contribution of the diagram in Fig. \ref{diagrams1} (right) with $\chi_R$ and $\nu_L$ running in the loop. 
	This cancellation is exact if the masses of $\chi$ and $\nu_H$ are equal. Otherwise, 
	the sum of both contributions is logarithmically divergent (see, for instance, 
	\cite{Perelstein:2003wd,Han:2005ru} for the analogous cancellation in the collective breaking case). 
	Hence, $\chi_R$ can not be ignored \cite{Low:2004xc}.~\footnote{Incomplete 
		$SO(5)$ fermion representations also result in 
		two--loop quartically divergent contributions to the Higgs mass induced 
		by the couplings of their kinetic term \cite{Cheng:2004yc}. 
		\label{twoloops}} 
	Few comments are in order.  The T--parity of $\chi_R$ is even in contrast with the other four 
	components in $\Psi_R$, which are odd. If we had chosen the T--transformed of $\Psi_R$ to be $- \Psi_R$ 
	and the T--parity of $\chi_R$ to be also odd, the lepton running in the diagram of Fig. \ref{diagrams1} (right) 
	would have been $\chi_R$ and $\nu_{H L}$, and not $\nu_L$. 
	Obviously, including all field components both in the LH $SU(5)$ and the RH $SO(5)$ 
	multiplets, the total contribution to the Higgs mass cancels due to the NG nature of the Higgs boson. 
	However, as we want only to duplicate the SM (LH) lepton doublets to start with, the SM singlet 
	$\chi_R$ must be always included to cancel the quadratically divergent contribution of $\nu_H$ to 
	the Higgs mass. 
	This can be checked diagramatically working out the corresponding Feynman rules for the 
	Lagrangian in Eq. (\ref{mirror}) and computing the 
	diagrams in Fig. \ref{diagrams1}, or reading the Higgs mass from the general Coleman-Weinberg 
	expression \cite{ArkaniHamed:2002qy,Coleman:1973jx,Ferrara:1994kg}
	\bea
	{\mathcal V}^{1-loop} = \frac{1}{32\pi^2} {\rm Str}( {\cal M}^2 ) \Lambda^2 + 
	\frac{1}{64\pi^2} {\rm Str} ( {\cal M}^4 ) 
	\ln \frac{{\rm Str}( {\cal M}^2)}{\Lambda^2} + \dots\ , 
	\eea
	where ${\rm Str}( {\cal M}^n ) = \sum_p (-1)^{2s_p} (2s_p+1) m_p^n$ runs over all particles 
	with spin $s_p$ and background dependent mass $m_p$ and 
	$\Lambda$ is the momentum cut--off $\sim 4 \pi f$.
	\footnote{Note that for fermions ${\cal M}^2 \equiv {\cal M} {\cal M}^\dagger$.}
	
	The lepton singlets $\chi_R$ must also get a large (vector--like) mass 
	by combining with a LH singlet $\chi_L$ 
	through a direct mass term without further couplings to the Higgs. 
	As they must do the extra leptons (partner doublets) in $SO(5)$ multiplets, ${\psi'}_R$ in Eq. (\ref{complete}), 
	for they must be also included in order to keep 
	the LFV Higgs decay amplitudes into charged leptons one--loop finite \cite{delAguila:2017ugt}. 
	(See also footnote \ref{twoloops}.) 
	Thus, their mass terms write 
	\bea
	{\cal L}_M = - M \overline{\chi_L} \chi_R  - M' \overline{{\psi'_L}} {\psi'_R} + {\rm h.c.} \ , 
	\label{direct}
	\eea
	where 
	${\psi}'_L$ is the LH (doublet) counterpart of ${\psi}'_R$. 
	\footnote{The Higgs boson mass is also free of quadratically divergent contributions of order 
		$\kappa^2$ if the $SO(5)$ (RH) multiplets are complete and the SM singlets $\chi_L$ are 
		doubled by including 
		them in the $SU(5)$ multiplets $\Psi_{1, 2}$ in Eq. (\ref{Psimultiplets}).
		However, the Yukawa Lagrangian in Eq. (\ref{mirror}) also provides a large mixing between 
		$\nu_L$ and (the T--even combination of) $\chi_L$, whose relatively large value $\sim v/2f$ is fixed. 
		Hence, new mass term contributions are needed to make the lepton singlets heavier (than 
		$\sim \sqrt{2} \kappa f$) and 
		satisfy the limit on the singlet content of light neutrinos (mainly electro--weak doublets), which is bound to be $< 0.03$ at 95 \% C.L.
		\cite{delAguila:2008pw,deBlas:2013gla,Thesis} (see below), 
		without pushing $f$ too high. We will not consider this enlarged lepton content any further.} 
	(Nevertheless, in the following we will not be concerned with the mirror leptons or their partners 
	${\psi}'$ because they are T--odd and do not mix with the SM leptons nor with $\chi$.) 
	With this matter content, $\chi_L$ is an $SU(5)$ singlet and it is therefore natural to include 
	a small Majorana mass for it. 
	Once LN is assumed to be only broken by {\tiny }small Majorana masses $\mu$ 
	in the heavy LH neutral sector, 
	\bea
	{\cal L}_\mu = - \frac{\mu}{2} \overline{\chi^c_L} \chi_L + {\rm h.c.} \ , 
	\label{Majorana}
	\eea
	the resulting (T--even) neutrino mass matrix reduces to the inverse see--saw one: 
	\bea
	{\cal L}^\nu_M = - \frac{1}{2} 
	\left( \begin{array}{ccc}
		\overline{\nu_L^c} &
		\overline{\chi_R} &
		\overline{\chi_L^c} 
	\end{array} \right) 
	{\cal M}^{T-even}_\nu 
	\left( \begin{array}{c}
		\nu_L\\
		\chi^c_R \\
		\chi_L 
	\end{array} \right) + {\rm h.c.} \ , 
	\label{seesaw}
	\eea
	where 
	\bea
	{\cal M}^{T-even}_\nu = 
	\left(  
	\begin{array}{ccc} 
		0 & i \kappa^* f \sin \big(\frac{v}{ \sqrt{2} f}\big) & 0 \\
		i \kappa^\dagger f \sin \big(\frac{v}{ \sqrt{2} f}\big) & 0 & M^\dagger \\
		0 & M^* & \mu 
	\end{array}
	\right) \ , 
	\label{Matrix}
	\eea
	with each entry standing for a $3 \times 3$ matrix to take into account the 3 lepton families. 
	The $\kappa$ entries are given by the Yukawa Lagrangian in Eq. (\ref{mirror}) and $M$ stands 
	for the direct heavy Dirac mass matrix in Eq. (\ref{direct}), while $\mu$ is the mass matrix of 
	small Majorana masses in Eq. (\ref{Majorana}). The natural size of the mass eigenvalues 
	for $M$ is $\sim 10\ {\rm TeV}$, of the order of $4 \pi f$ with $f \sim {\rm TeV}$,  
	as required by current EWPD (see below) if we assume the $\kappa$ eigenvalues to be order 1. 
	While the $\mu$ eigenvalues shall be much smaller than the GeV. 
	The predictions for the SM neutrino masses and the LFV contributions of the 
	quasi--Dirac singlets $\chi$ are those of the inverse see--saw  \cite{Arganda:2014dta,DeRomeri:2016gum,Ballett:2019bgd}. 
	(See \cite{Malinsky:2005bi,Agashe:2016ttz,Agashe:2017ann,DeRomeri:2018pgm,Agashe:2018cuf,CarcamoHernandez:2019eme} 
	for analyses in alternative SM extensions, including models with warped extra dimensions.) 
	Before going through the corresponding phenomenological study, 
	let us comment on two other a priori less natural scenarios.  
	
	\subsection{T-odd heavy singlet}
	\label{Tparity}
	
\begin{figure}
		\centering
		\begin{tabular}{ccc}
			\includegraphics[scale=0.35]{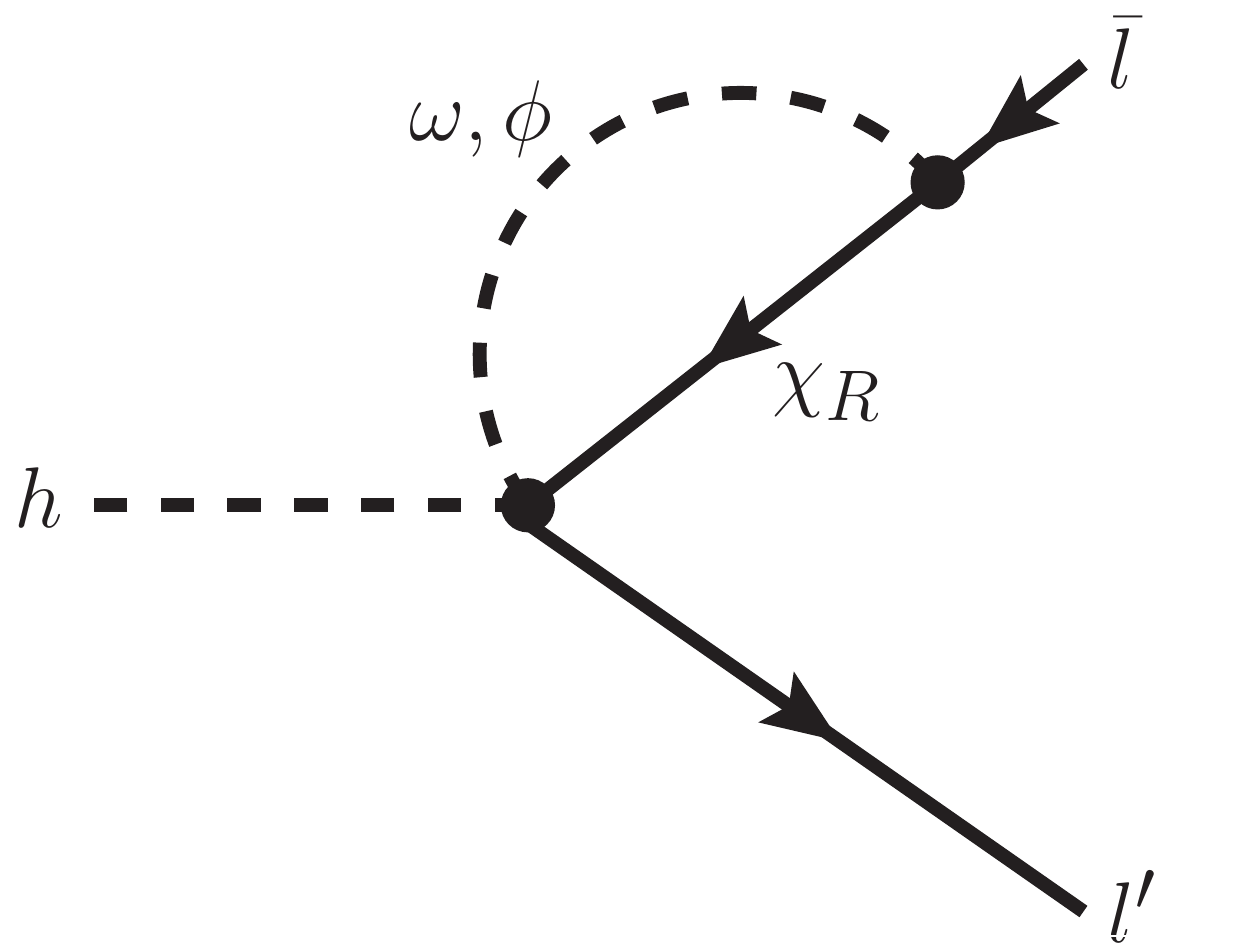} \quad & 
			\quad $\begin{array}{c}  \\[-1.8cm] + \\[1.8cm] \end{array}$ \quad & \quad 
			\includegraphics[scale=0.35]{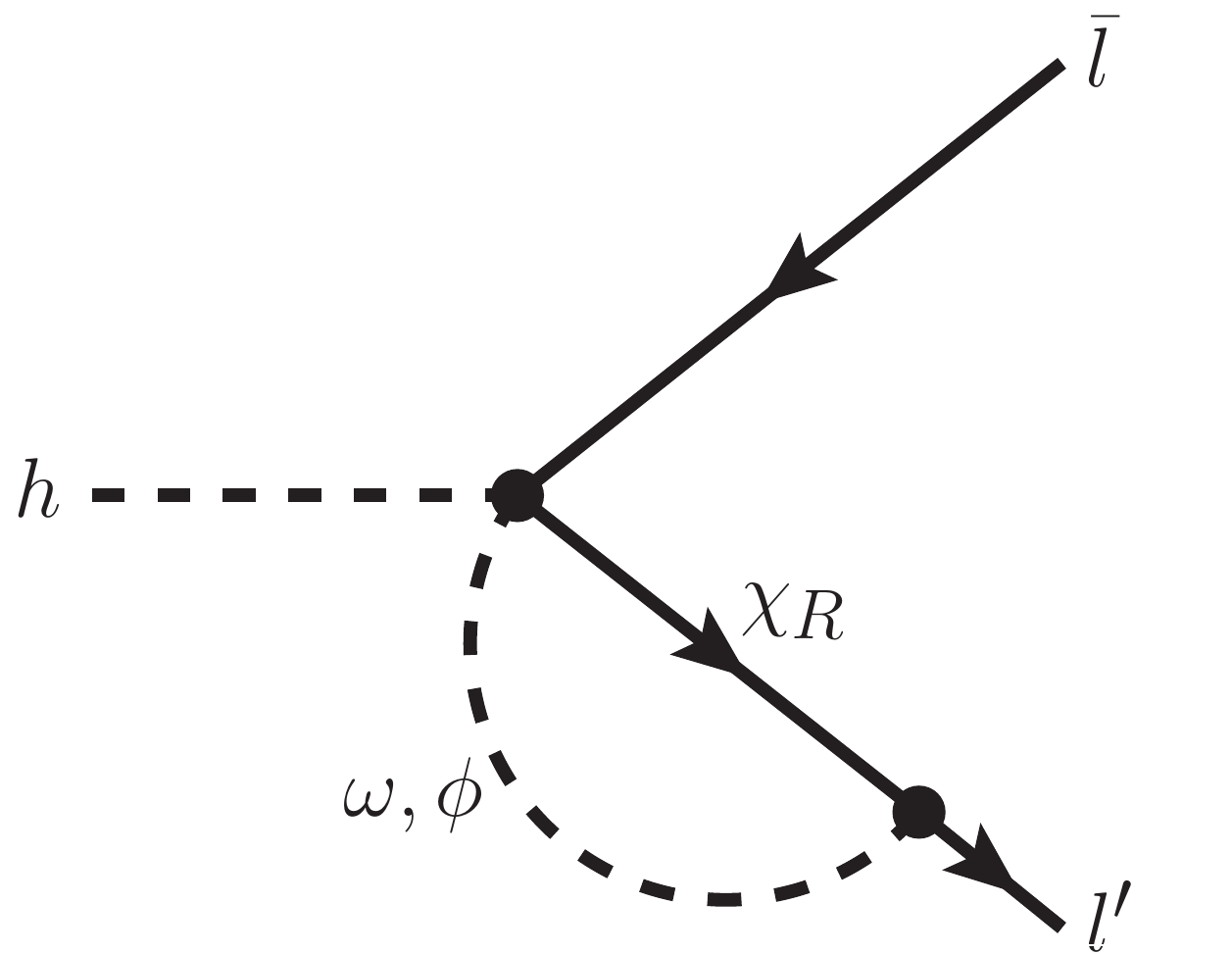} 
		\end{tabular}
		\caption{Higgs decay diagrams exchanging the RH singlet in $SO(5)$ quintuplets 
			resulting in an unmatched divergent contribution when SM charged leptons get their 
			mass from the usual (minimal) Yukawa coupling.}
		\label{diagrams2}
	\end{figure}

	If we had chosen the T--parity of $\chi_R$ to be odd by 
	defining the T--action on the fermions 
	$\Psi_1\stackrel{{\rm T}}{\longleftrightarrow}-\Sigma_0\Psi_2$, 
	$\Psi_R\stackrel{{\rm T}}{\longrightarrow}-\Psi_R$
	and hence, the T--invariant Yukawa Lagrangian in Eq. (\ref{mirror}) to be 
	$\lag_{Y_H} = -\kappa f 
	\left(\overline\Psi_2\xi+ \overline\Psi_1\Sigma_0
	\Omega\xi^\dagger\Omega\right)\Psi_R + {\rm h.c.}$, 
	all new fermions would be T--odd 
	\cite{Low:2004xc}. 
	Thus, their contribution to the mass of the SM neutrinos,  
	once LN is broken as assumed before, would be one--loop suppressed. 
	This appealing possibility has the drawback that the LFV Higgs decays 
	into two charged leptons become logarithmically divergent due to the 
	contribution of $\chi_R$ when exchanged in the diagrams in Fig. \ref{diagrams2}. 
	(We will provide further details elsewhere.) 
	As we are interested in setting up a predictive model at leading order at least to one loop, 
	meaning in our case that the Higgs boson mass can get one--loop corrections 
	at most logarithmically divergent and that LFV processes involving only SM external 
	fields must be finite, we disregard this alternative in what follows. 
	Nevertheless, the logarithmically divergent contribution of $\chi_R$ 
	to LFV Higgs decays into charged fermions does rely on the mechanism 
	giving masses to them. 
	In our case, 
	we assume that the charged leptons $\ell$ get their masses through the 
	Yukawa Lagrangian 
	\cite{Hubisz:2004ft,Chen:2006cs,Goto:2010sn} 
	(summation over $x, y, z = 3, 4, 5$ and $r, s = 1, 2$ is understood): 
	\bea
	{\cal L}_Y = \frac{i \lambda}{2 \sqrt 2} f \epsilon^{xyz}\epsilon^{rs} 
	\left[ (\overline{\Psi_2^\sigma})_x (\Sigma)_{ry} (\Sigma)_{sz} +  
	(\overline{\Psi_1^{\sigma'}} \Sigma_0 \Omega)_x ({\Sigma'})_{ry} ({\Sigma'})_{sz} \right] 
	\ell_R +{\rm h.c.}\ , 
	\label{SMmasses}
	\eea
	where the LH leptons are included in two other 
	incomplete $SU(5)$ multiplets in fundamental representations $\underline{5}$ and 
	$\underline{5}^*$, respectively: 
	\bea
	\Psi^{{\sigma'}}_1=\left(\ba{c} {\sigma'} l_{1L} \\ 0 \\ 0 \ea\right),\quad
	\Psi^\sigma_2=\left(\ba{c} 0 \\ 0 \\ \sigma l_{2L} \ea\right) , 
	\label{Psichimultiplets}
	\eea
	with $\sigma, {\sigma'}$ scalars 
	with the proper charges to endow $\sigma l_{2L}$ and ${\sigma'} l_{1L}$ with the charges 
	of the corresponding components of $\underline{5}^*$ and $\underline{5}$, respectively, 
	and $\Sigma' = \Omega \Sigma_0 \Sigma^\dagger \Sigma_0 \Omega$. 
	Thus, the introduction of the scalars $\sigma, {\sigma'}$ allows us to change the sign of the 
	gauged $U(1)$ charges in 
	$SU(5)$ for $l_{2L}$ and $l_{1L}$ while also giving the correct hypercharge to $\sigma l_{2L}$ and ${\sigma'} l_{1L}$. 
	The action under T--parity is then defined as
	\bea
	\Psi^{{\sigma'}}_1\stackrel{{\rm T}}{\longleftrightarrow}\Omega\Sigma_0\Psi^\sigma_2\ .
	\label{PsichiT}
	\eea
	However, this particular construction does not allow to allocate $\chi_L$ 
	in $\Psi^{{\sigma'}}_1$ or $\Psi^\sigma_2$ and then, 
	no coupling to $\ell_R$ can compensate for 
	the logarithmically divergent contribution of $\chi_R$ in Fig. \ref{diagrams2}. 
	If we wanted to insist in $\chi$ being T--odd and hence, in 
	introducing the adequate $(h+v)^2\omega^+\overline{\chi_L} \ell_R$ and 
	$(h+v)^2\Phi^+\overline{\chi_L} \ell_R$ couplings 
	to compensate this $\chi_R$ contribution, 
	we would have to assign $\ell_R$ to a larger representation, for instance, 
	generalizing the proposal for composite Higgs models 
	advocated in \cite{Contino:2006qr,Csaki:2008zd}, as we will review elsewhere. 
	
	\subsection{See--saw of type I and II}
	\label{typeII}
	
	As mentioned when describing the NG boson content of the model at the beginning 
	of this section, the triplet $\Phi$ has the correct SM 
	quantum numbers to mediate the see--saw of type II. 
	What originally brought to consider this mechanism to generate neutrino masses 
	in the LHT \cite{Han:2005nk}. 
	However, the contribution of a non--zero vev for $\Phi^0$ is expected to be subleading, 
	as we shall argue. 
	The Yukawa couplings giving large masses to mirror leptons in Eq. (\ref{mirror}) 
	fix the $\Phi$ LN to be zero and quantum corrections do not generate the see--saw operator of type II 
	\cite{Schechter:1980gr,Magg:1980ut,Cheng:1980qt,Gelmini:1980re,Lazarides:1980nt,Mohapatra:1980yp}:
	\bea
	{\mathcal L}_{see-saw}^{II} =  y_{ij}  {\overline{\tilde l_{L i}}}
	\,\Delta\, l_{L j} + {\rm h.c.} 
	\rightarrow - \frac{1}{2} m_{\nu j i}^{*} {\overline{\nu_{L i}^c}} \nu_{L j} + {\rm h.c.} , 
	\; {\rm with}\; m_{\nu j i}^* = \sqrt{2} y_{ij} \langle \Phi^0 \rangle\ , 
	\label{seesawII}
	\eea
	where $\tilde l_{L i} = i \sigma^2 l_{L i}^c = \left( \ell_{L i}^c \; -\nu_{L i}^c \right)^T$, 
	with $l_{L i}^c$ the three charge--conjugated SM lepton doublets, $i = 1, 2, 3$, and 
	\bea
	\Delta = \left( \begin{array}{cc} \dis\frac{\Phi^+}{\sqrt{2}} & -  \Phi^{++} \\ 
		\dis\frac{\Phi^0 + i \Phi^P}{\sqrt{2}} & - \dis\frac{\Phi^+}{\sqrt{2}} \end{array}\right)\ . 
	\eea
	(Note that we have included a $-i\sigma^2$ factor on the right in the 
	definition of $\Delta$ to take care of this factor in the definition of $\Psi$ in Eq. (\ref{Psimultiplets}).) 
	This coupling not only violates LN, which must be assumed to be broken at 
	some stage, but also T--parity because $\Phi$ is T--odd while the SM 
	fermions are T--even. 
	Obviously, T--parity is spontaneously broken if $\langle \Phi^0 \rangle \neq 0$ 
	and SM neutrinos shall get a mass once LN is broken. 
	Nevertheless, these masses must be induced by a T--parity preserving 
	operator, which must then involve an even number of $\Phi^0$'s and be 
	of higher dimension than the see--saw operator of type II above. 
	As a matter of fact, an SM invariant operator must also involve at least two Higgs doublets 
	because they must compensate 
	for the hypercharge of the two lepton doublets. In summary, an SM and T invariant 
	LN violating operator involving $\Phi^0$ is at least suppressed by a factor 
	$\langle \Phi^0 \rangle ^2 / f^2$ relative to the SM Weinberg operator \cite{Weinberg:1979sa}, 
	as we show below for the inverse see--saw, 
	and then, it is subleading in the LHT.  
	
\begin{figure}
		\centering
		\begin{tabular}{ccc}
			\quad\quad \includegraphics[scale=0.3]{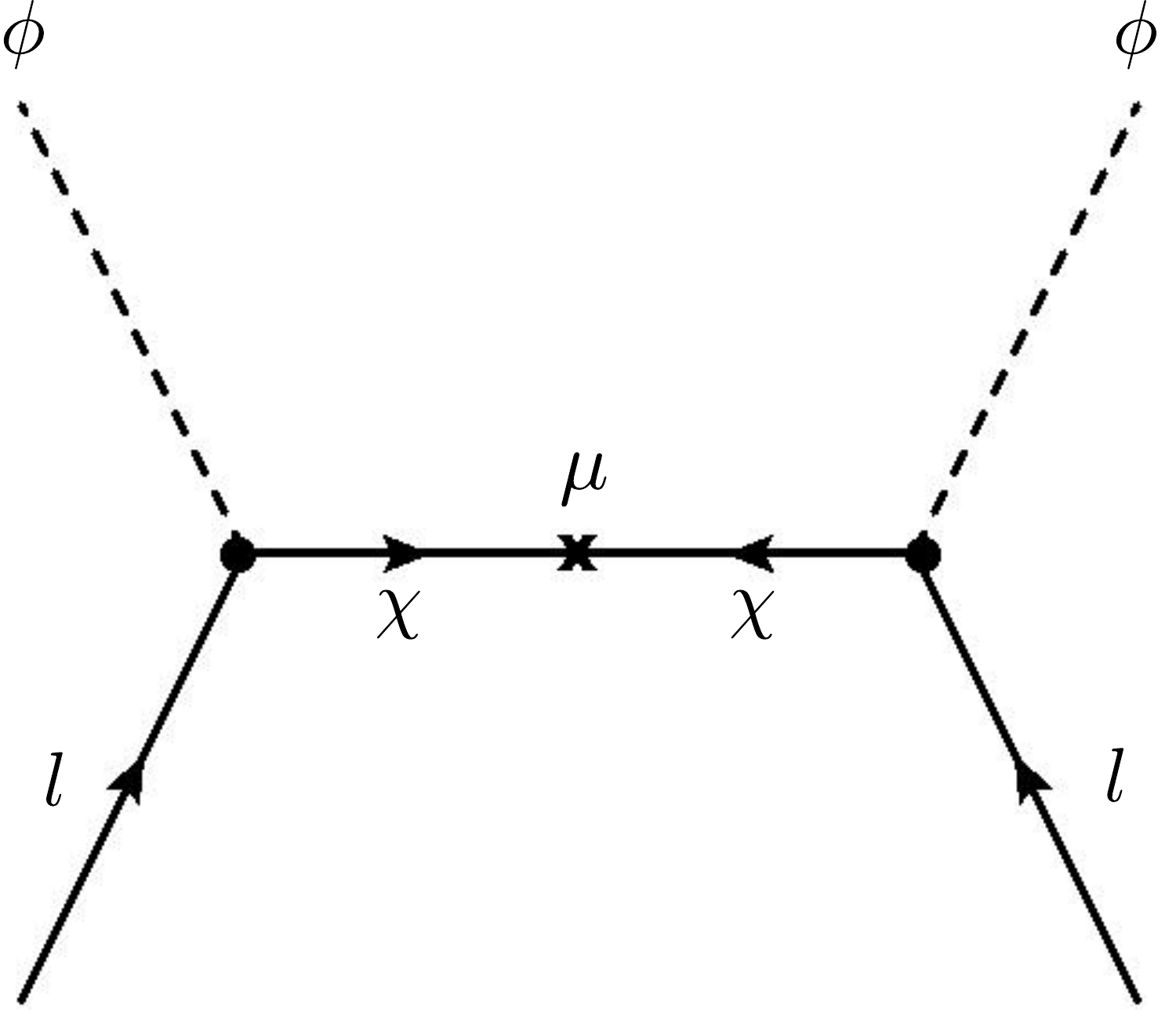} \quad & 
			\quad $\begin{array}{c}  \\[-2.3cm] + \\[2.3cm] \end{array}$ \quad & \quad
			\includegraphics[scale=0.3]{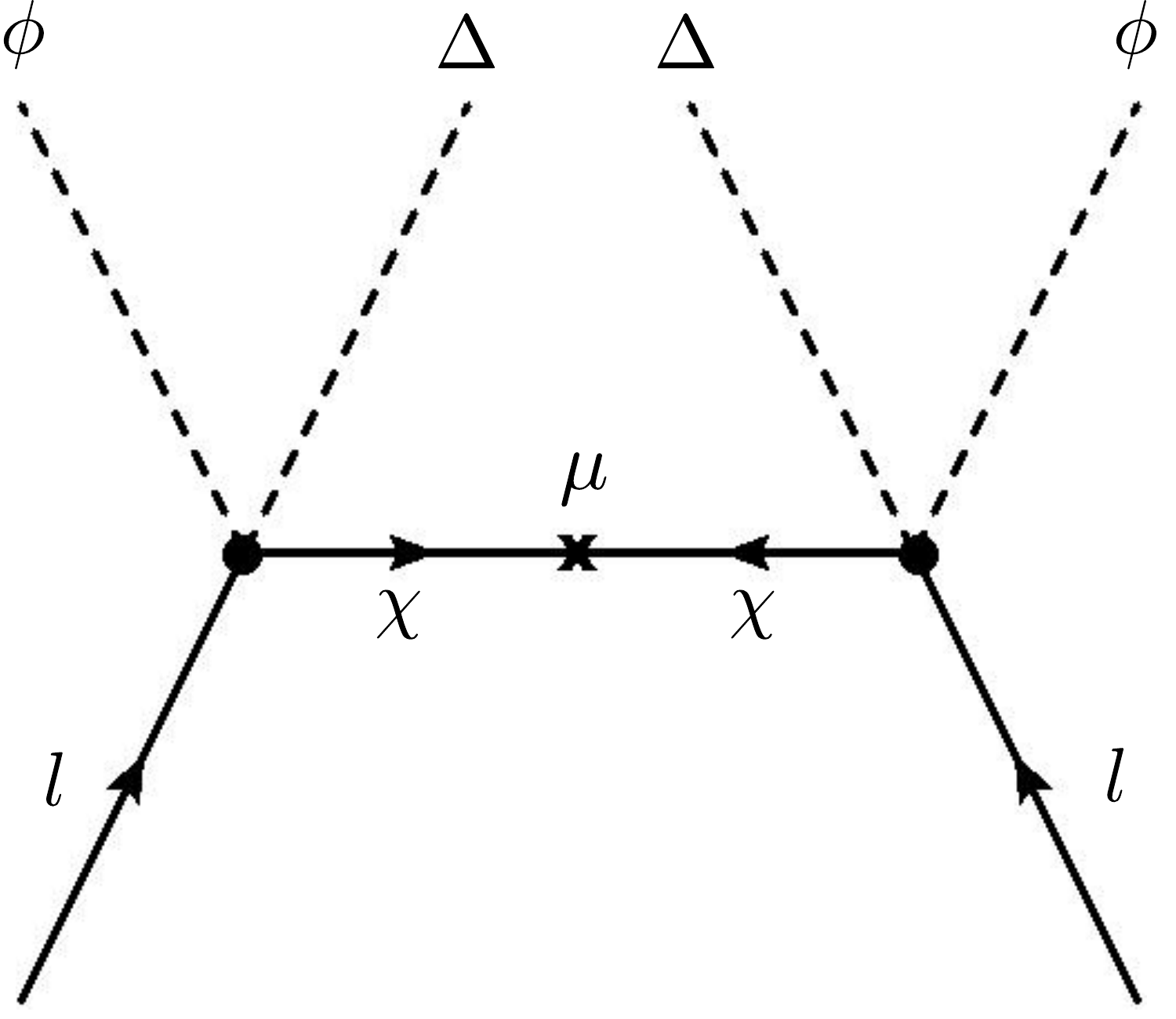} \\ 
			$\begin{array}{c}  \\[-2.3cm] + \\[2.3cm] \end{array}$ \quad \includegraphics[scale=0.3]{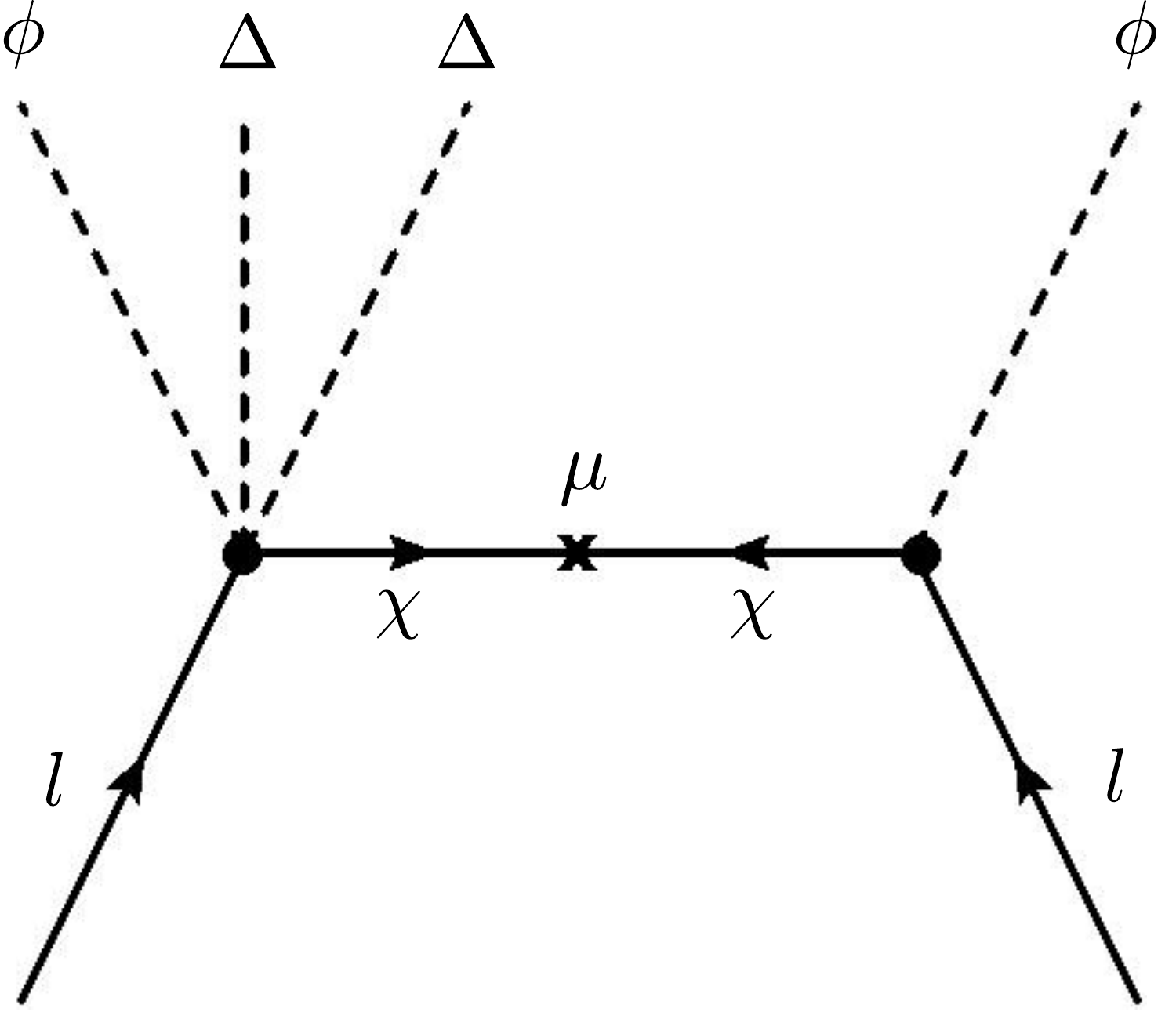} \quad & 
			\quad $\begin{array}{c}  \\[-2.3cm] + \\[2.3cm] \end{array}$ \quad & \quad
			\includegraphics[scale=0.3]{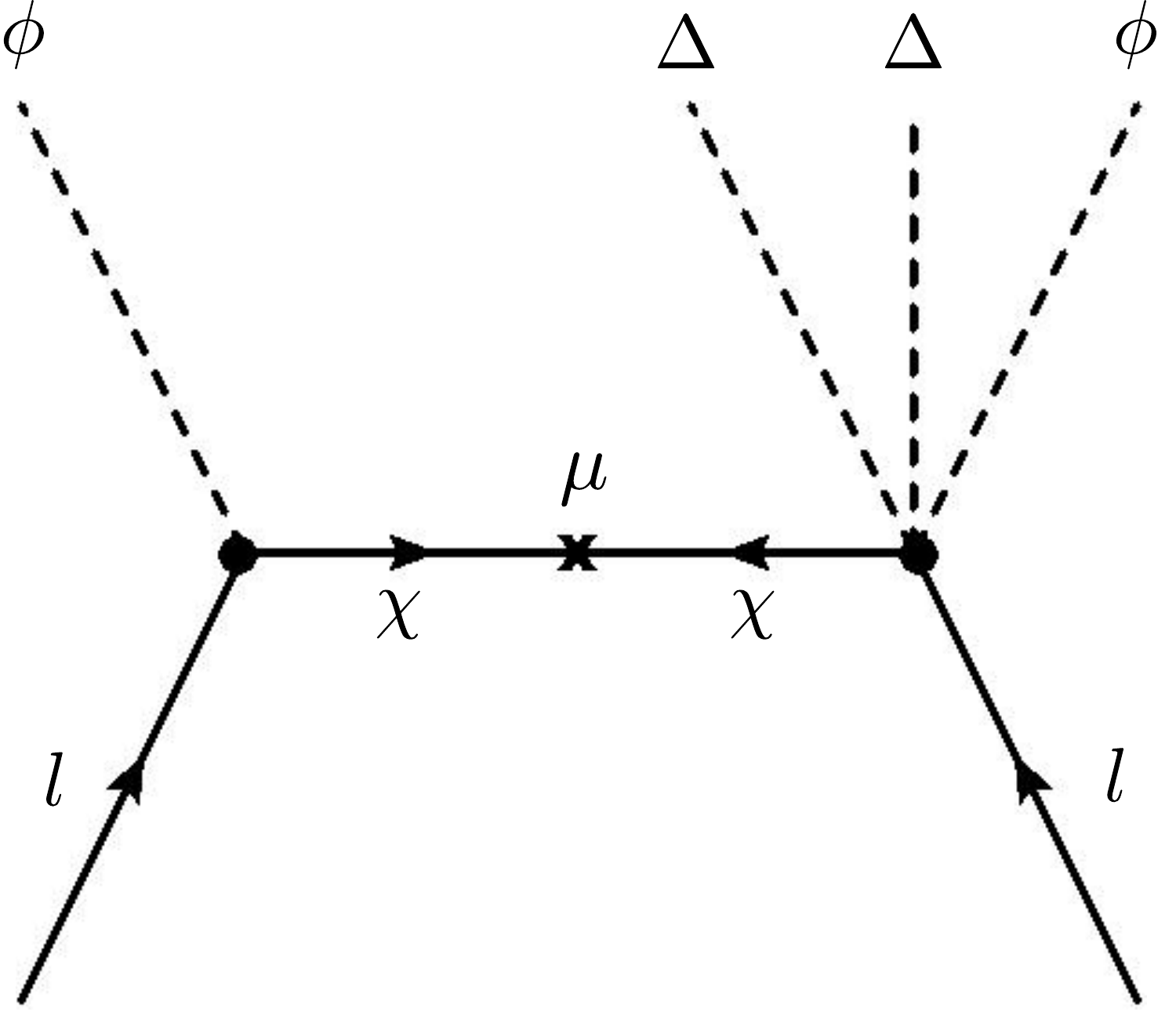} 
		\end{tabular}
		\caption{Diagrammatic expansion of the tree--level integration out of $\chi$, Eq. (\ref{pSMWO}).}
		\label{diagrams3}
	\end{figure}

	In the inverse see--saw model at hand the integration out of the 
	quasi-Dirac neutrinos $\chi$, with heavy masses given 
	to leading order by the $3\times3$ mass matrix $M$, generates the corresponding 
	Weinberg operator (see Eqs. (\ref{mirror}), (\ref{direct}) and (\ref{Majorana}))
	\footnote{The fermionic kinetic terms properly normalized are written, for instance, 
		in Ref. \cite{delAguila:2008zu}.}
	\bea
	\frac{1}{2} (\kappa f M^{-1})^* \mu (\kappa f M^{-1})^\dagger\ \left({\cal O}_\chi + {\cal O}'_\chi\right)\ , 
	\label{iWO}
	\eea
	with (also omitting family indices)
	\begin{align}
		{\cal O}_\chi = &  \overline{\Psi_1^c} \xi \frac{{\bf 1}+\Omega}{2} \xi^T \Psi_1 + 
		\overline{\Psi_2^c}  \xi^*  \frac{{\bf 1}+\Omega}{2} \xi^\dagger \Psi_2 \ ,\nonumber \\
		{\cal O}'_\chi = &  \overline{\Psi_1^c} \xi \frac{{\bf 1}+\Omega}{2} \xi^\dagger \Psi_2 + 
		\overline{\Psi_2^c}  \xi^* \frac{{\bf 1}+\Omega}{2} \xi^T \Psi_1  \ .
		\label{pWO}
	\end{align}
	Both ${\cal O}_\chi$ and ${\cal O}'_\chi$ include the SM Weinberg operator and subleading contributions 
	containing $\Phi^0$ (we omit subleading terms not involving $\Delta$)
	\begin{align}
		{\cal O}_\chi \supset & - \frac{1}{2 f^2}(\overline{l_L^c} \tilde{\phi}^*)(\tilde{\phi}^\dagger l_L) 
		- \frac{1}{4 f^4} \left[\,
		\frac{1}{2}(\overline{l_L^c} \Delta^T {\phi}^*)({\phi}^\dagger \Delta l_L) \quad \quad \;
		\right.  \nonumber \\
		& \left. -\, \frac{1}{3}(\overline{l_L^c} \Delta^T \Delta ^* \tilde{\phi}^* )(\tilde{\phi}^\dagger l_L) 
		- \frac{1}{3}(\overline{l_L^c} \tilde{\phi}^* )(\tilde{\phi}^\dagger \Delta^\dagger \Delta l_L)  \right] + \dots  \ ,
		\nonumber \\
		{\cal O}'_\chi \supset & - \frac{1}{2 f^2}(\overline{l_L^c} \tilde{\phi}^*)(\tilde{\phi}^\dagger l_L) 
		+ \frac{1}{4 f^4} \left[\,
		\frac{1}{2}(\overline{l_L^c} \Delta^T {\phi}^*)({\phi}^\dagger \Delta l_L) \quad \quad \;
		\right.  \nonumber \\
		& \left. +\, \frac{1}{3}(\overline{l_L^c} \Delta^T \Delta ^* \tilde{\phi}^* )(\tilde{\phi}^\dagger l_L) 
		+ \frac{1}{3}(\overline{l_L^c} \tilde{\phi}^* )(\tilde{\phi}^\dagger \Delta^\dagger \Delta l_L)   \right] + \dots  \ ,
		\label{pSMWO}
	\end{align}
	with $\tilde{\phi} = i \sigma^2 \phi^*$. 
	(The four terms in the operator expansion can be also read from the 
	diagrammatic tree--level integration out of $\chi$ in Fig. \ref{diagrams3}.)  
	As emphasized above, the dimension 7 operators give an extra contribution to the SM neutrino masses 
	but proportional to $\langle \Phi^0 \rangle ^2 v^2 / f^3$ and then subleading, being its ratio 
	to the leading term $-\langle \Phi^0 \rangle ^2 / 6 f^2$.
	\footnote{ 
		The corresponding LHT Weinberg operators preserving the full global symmetry write 
		(see also \cite{Blanke:2007db})  
		\bea
		{\cal O} = \overline{\Psi_1^c} \Sigma \Psi_1 + 
		\overline{\Psi_2^c} \Sigma^\dagger \Psi_2\ , \quad 
		{\cal O}' = \overline{\Psi_1^c} \Psi_2 + 
		\overline{\Psi_2^c} \Psi_1  \ , 
		\label{WO}
		\eea
		where $\Sigma = \xi \Sigma_0 \xi^T$ is a 
		$5\times 5$ symmetric tensor under $SU(5)$, $\Sigma {\rightarrow} V \Sigma V^T$, and 
		$\Sigma_0$, introduced in Eq. (\ref{TPsi}), is the singlet direction under $SO(5)$, $\Sigma_0 = U \Sigma_0 U^T$. 
		The expansion of ${\cal O}$ also includes the SM Weinberg operator  
		as well as the lowest order operators (of dimension 7) involving the scalar triplet $\Phi$,  
		\bea
		{\cal O} \supset - \frac{1}{f^2}(\overline{l_L^c} \tilde{\phi}^*)(\tilde{\phi}^\dagger l_L) 
		- \frac{1}{3 f^4} 
		\left[ \,(\overline{l_L^c} \Delta^T {\phi}^*)({\phi}^\dagger \Delta l_L) \right. \quad \quad \;
		\nonumber \\ 
		- (\overline{l_L^c} \Delta^T \Delta ^* \tilde{\phi}^* )(\tilde{\phi}^\dagger l_L) 
		- (\overline{l_L^c} \tilde{\phi}^* )(\tilde{\phi}^\dagger \Delta^\dagger \Delta l_L) \left. \right] + \dots  \ .
		\label{SMWO}
		\eea
		While ${\cal O}'$ has no scalar couplings, and with the fermion content in Eq. (\ref{Psimultiplets}) it vanishes.}
	Thus, the scalar triplet contribution to neutrino masses is expected to 
	be very much suppressed relative to the see-saw of type I.
	(We shall assume in the following a vanishing $\langle \Phi^0 \rangle$.) 
	
	\section{Inverse see--saw masses and mixings}
	\label{Constraints}
	
	The inverse see--saw has been widely studied in the literature 
	\cite{Abada:2007ux,Arganda:2014dta,DeRomeri:2016gum,Ballett:2019bgd,Malinsky:2005bi,DeRomeri:2018pgm,CarcamoHernandez:2019eme}. 
	The light ($l$) neutrino masses can be obtained diagonalizing the neutrino mass 
	matrix in Eq. (\ref{Matrix}) to leading order 
	or directly from Eqs. (\ref{iWO}) and (\ref{pSMWO}): 
	\bea
	({\cal M}^l_\nu)_{i j} = \theta^*_{i k} \mu_{k l} \theta^*_{j l}\ , \quad {\rm with} \quad 
	\theta_{i k} = - i f \sin \Big(\frac{v}{\sqrt{2} f}\Big) \kappa_{i k} M^{-1}_k \ ,
	\label{lightmasses}
	\eea
	where we have reintroduced the family indices 
	(summation is understood when they are repeated in a product) 
	and assumed without lost 
	of generality that the $\chi$ mass matrix, $M$, is diagonal and positive definite. 
	At the same time (in the basis where the charged lepton mass matrix is diagonal)
	\bea
	{\cal M}^l_\nu = U_{\rm PMNS}^* {\cal D}^l_\nu U_{\rm PMNS}^\dagger \; 
	{\rm and\ solving\ Eq.\ (\ref{lightmasses}) ,} \; 
	\mu = (\theta^*)^{-1} U_{\rm PMNS}^* {\cal D}^l_\nu U_{\rm PMNS}^\dagger (\theta^\dagger)^{-1} ,
	\label{dlightmasses}
	\eea
	where $U_{\rm PMNS}$ is the Pontecorvo--Maki--Nakagawa--Sakata 
	mixing matrix 
	\cite{Pontecorvo:1957cp,Pontecorvo:1957qd,Pontecorvo:1967fh,Maki:1962mu} 
	and ${\cal D}^l_\nu$ the diagonal neutrino mass matrix. 
	Hence, for a non--singular $\theta$ matrix $\mu$ can be always adjusted to fit light neutrino 
	masses and mixing. 
	(We use $\mu$ to denote the Majorana matrix for $\chi_{L i}$ or any of its small entries, 
	what should be clear by the context.) 
	For instance, if $f > 1$ TeV, $\kappa = 1$ and $M = 10$ TeV, 
	$\mu \sim 0.3$ keV for a light neutrino mass of 0.1
        eV.~\footnote{Note that the small mass parameters $\mu\sim0.3
          ~\textrm{keV}$ are technically
          natural~\cite{tHooft:1979rat}, as they are the only terms in
          the theory breaking LN. Although in the absence of a flavor symmetry
          there is no dynamical explanation for the $\kappa$ and $\mu$ values fulfilling Eq. (\ref{dlightmasses}).}  
	
	The experimental limits on $\theta$ can be always satisfied without implementing flavor symmetries in the model 
	but LFV constraints set stringent limits 
	on the heavy scale as well as on the mixing between light and heavy leptons,  
	as we review in the following. 
	
	\subsection{LFV limits}
	\label{LFV}
	
	The $\theta$ matrix elements give the mixing between light and heavy 
	(quasi--Dirac) neutrinos, $l$ and $h$, respectively,  
	\begin{equation} 
		(U_{\rm PMNS})_{i j} \nu_{L j}^{l} = [{\bf 1}_{3\times3} - \frac{1}{2} (\theta \theta^\dagger)]_{i j} 
		\nu_{L j} - \theta_{i j} \chi_{L j}\ ,\;\;
		\chi_{L i}^{h} = [{\bf 1}_{3\times3} - \frac{1}{2} (\theta^\dagger \theta)]_{i j}
		\chi_{L j} + \theta^\dagger_{i j} \nu_{L j}\ ,
		\label{mixing}
	\end{equation}
	to leading order. 
	They are constrained by lepton flavor conserving processes 
	at tree level because they modify the SM charged and neutral currents 
	(in standard notation) \cite{delAguila:2000aa,delAguila:2000rc}:~
	\footnote{Charged and neutral currents are related at leading order $X_{i j} = W_{i k} W^*_{j k}$, 
		with the neutral currents satisfying the positivity constraints 
		$|X_{i j}|^2 \leq X_{i i} X_{j  j}$, $|\delta_{i j} - X_{i j}|^2 \leq (1 - X_{i i}) (1 - X_{j  j})$. 
		The latter reduces to the Schwarz inequality 
		$|(\theta \theta^\dagger)_{i j}|^2 \leq (\theta \theta^\dagger)_{i i} (\theta \theta^\dagger)_{j j}$. 
		All of them are automatically taken care working only with the mixing matrix elements $\theta_{i j}$.} 
	\bea
	{\cal L}_W^l &=&  \frac{g}{\sqrt{2}} \overline{\nu^l_{L i}} {W}_{i j} \gamma^\mu \ell_{L j} W^+_\mu 
	+ {\rm h.c.}\ , 
	\quad {\rm with} \quad  {W}_{i j} = \{U^\dagger_{\rm PMNS} [{\bf 1}_{3\times3} - \frac{1}{2} (\theta \theta^\dagger)]\}_{i j} \ , \nonumber \\ 
	{\cal L}_Z^l &=&  \frac{g}{2 c_W} \overline{\nu^l_{L i}} {X}_{i j} \gamma^\mu \nu^l_{L j} Z_\mu\ , 
	\quad {\rm with} \quad  {X}_{i j} = 
	\{U^\dagger_{\rm PMNS} [{\bf 1}_{3\times3} - (\theta \theta^\dagger)] U_{\rm PMNS}\}_{i j} \ .
	\label{SMcurrents}
	\eea 
	More stringent are the constraints from (charged) LFV processes which  
	proceed at one loop, as do $(g-2)_\ell$ and at higher order the 
	Electric Dipole Moment of the electron (EDM$_e$).
	\footnote{The addition of heavy neutrinos does not modify the SM neutral 
		currents for charged leptons at tree level and then, 
		they remain lepton flavor conserving and universal.} 
	Even though they are suppressed by the corresponding loop factors $1/16 \pi^2$, 
	they can and do significantly restrict the $\theta$ matrix elements (and the heavy neutrino masses $M_i$) 
	fixing the coupling between the SM leptons and the heavy 
	quasi--Dirac neutrinos:  
	\bea
	{\cal L}_W^{lh} =  \frac{g}{\sqrt{2}} 
	\overline{\chi^h_{L i}} {\theta}^\dagger_{i j} \gamma^\mu \ell_{L j} W^+_\mu + {\rm h.c.}\ , \quad   
	{\cal L}_Z^{lh} =  \frac{g}{2 c_W} 
	\overline{\chi^h_{L i}} (\theta^\dagger U_{\rm PMNS})_{i j} \gamma^\mu \nu^l_{L j} Z_\mu + {\rm h.c.}\ .
	\label{lhchargedcurrents}
	\eea
	The Yukawa coupling in Eq. (\ref{mirror}) also enters in the calculation of Higgs decays, 
	for instance, 
	\bea
	{\cal L}_h^{\nu} \supset \frac{i}{\sqrt{2}} \cos\left(\frac{v}{\sqrt{2} f}\right) 
	\overline{\nu_{L i}} \kappa_{i j} \chi_{R j} h + {\rm h.c.} \simeq 
	- \overline{\nu^l_{L i}} (U^\dagger_{\rm PMNS} \theta)_{i j} \frac{M_j}{v} \chi_{R j} h 
	+ {\rm h.c.}\ ,  
	\label{lhcurrents}
	\eea
	where the last equation gives the leading term in $v/f$ and $\theta_{i j}$.
	
	In order to properly confront the LHT with experiment we should perform a global fit to 
	EWPD and to current LFV experimental limits. 
	This is, however, beyond the scope of this paper, in particular because there are also 
	other one--loop contributions to the latter mediated by T--odd leptons 
	\cite{delAguila:2017ugt,delAguila:2019htj}. 
	Moreover, while the amplitudes exchanging T--odd leptons are suppressed by 
	inverse powers of $f$, the amplitudes exchanging heavy T--even neutrinos are 
	suppressed by inverse powers of their masses $M_k$ and hence, their sizes can be 
	made to vary a priori independently. 
	We would then only derive conservative bounds, postponing a global fit to a future publication. 
	
	In the top part of Table \ref{Limits} we collect the limits from EWPD obtained 
	assuming that each heavy neutrino only mixes with one light neutrino of 
	definite flavor and that only one mixing is non--vanishing at a time 
	\cite{delAguila:2008pw,deBlas:2013gla,Thesis}. 
	\begin{table}
		\begin{center}
					\hspace*{-1cm}
			\begin{tabular}{|c|c|c|}\hline
				\multicolumn{3}{|c|}{ EWPD\quad (only one $\theta_{i i} \neq 0$, at 95 \% C.L.\; \cite{deBlas:2013gla}) }\\\hline
				$| \theta_{e 1} | < 0.04$ & 
				$| \theta_{\mu 2} | < 0.03$ &
				$| \theta_{\tau 3} | < 0.09$ \\\hline\hline
				\multicolumn{3}{|c|}{ LFV\quad at 90 \% C.L.\; ($M_k = 10$ TeV) }\\\hline
				Br($\mu \to e\ \gamma$) < $4.2 \times 10^{-13}$ \cite{Adam:2013mnn} & 
				Br($\tau \to e\ \gamma$) < $3.3 \times 10^{-8}$ \cite{Aubert:2009ag} &
				Br($\tau \to \mu\ \gamma$) < $4.4 \times 10^{-8}$ \cite{Aubert:2009ag} \\ 
				| $\theta_{e j} \theta_{\mu j}^* | < 0.14 \times 10^{-4}$ & 
				| $\theta_{e j} \theta_{\tau j}^* | < 0.40 \times 10^{-2}$ & 
				| $\theta_{\mu j} \theta_{\tau j}^* | < 0.46 \times 10^{-2}$ \\\hline
			\end{tabular}
			\caption{Limits on the mixing between the SM and the heavy quasi--Dirac neutrinos from 
				electro--weak precision data (top) and from lepton flavor violating processes (bottom). 
				The sum on the repeated index $j = 1, 2, 3$ is understood.} 
			\label{Limits}
		\end{center}
	\end{table}
	This means that only $\theta_{i i} \neq 0$ in the basis where the charged leptons are diagonal. 
	Assuming universality and, in particular, that the three mixings $\theta_{i i}$ are equal, 
	their absolute value is found to be $< 0.03$ at 95 \% C.L. \cite{Thesis}. 
	Hence, Eq. (\ref{lightmasses}) implies 
	\bea
	|\kappa_{i i}|\ < \ 0.17\ \left(\frac{M_i}{{\rm TeV}}\right)\ , 
	\label{treebound}
	\eea
	for $f$ larger than the TeV. 
	Note that this effective description requires $M_i \lesssim 4 \pi f \sim 10$ TeV for 
	consistency of the model. 
	What translates into an upper bound on $\kappa_{i i}$ and in turn, into an upper 
	bound on the mass of (T--odd) mirror leptons $\simeq \sqrt{2} \kappa f$ 
	(see below). 
	
	LFV further restricts the mixing between light and heavy neutrinos, 
	especially for the first two families. 
	A solution satisfying current bounds is to assume $\theta$ diagonal, as above, 
	banishing LFV for none has been observed up to now. 
	However, the Yukawa coupling $\kappa$ is an arbitrary $3\times 3$ matrix and 
	hence, in general 
	\bea
	\kappa = V^\dagger {\kappa}^{diag} Z \ ,  
	\label{treebound}
	\eea
	with $V$ and $Z$ unitary matrices and ${\kappa}^{diag}$ a diagonal matrix 
	with semipositive eigenvalues. 
	$V$ is the transformation matrix 
	relating the mass eigenvector basis for $l_{H L}$ with the $\ell_L$ one \cite{delAguila:2008zu} 
	and $Z$ is the transformation relating the mass eigenvector basis for $l_{H R}$ with the $\chi_R$ one. 
	Nevertheless, this parameterization will only matter when performing a 
	general global fit. 
	When performing it we shall find that for particular 
	values of these Yukawa couplings some of the LFV observables can cancel, as found  
	when studying the contributions of the T--odd leptons in \cite{delAguila:2017ugt,delAguila:2019htj}. 
	But not all of them will vanish at the same time, except in the singular case when all heavy leptons are 
	degenerate or the heavy sector is aligned with the SM. The allowed parameter region 
	will be then restricted by the non--vanishing observables. 
	The size of this region and the amount of fine--tuning are determined 
	by the most stringent bounds. 
	However, such a phenomenological discussion is beyond the scope of this paper, 
	as already emphasized. 
	In order to estimate the size of these regions is sufficient to consider 
	the most restrictive current bounds, which are obtained from the non--observation 
	of the radiative decays $\ell \to \ell' \gamma$ (see Table 11 in \cite{delAguila:2019htj}, 
	and Ref. \cite{Baldini:2018uhj}). 
	In Table \ref{Limits} we gather the corresponding limits. 
	The contribution of the heavy (quasi--Dirac) neutrinos can be evaluated 
	in the 't Hooft--Feynman gauge 
	in a similar way as the contribution of the (T--odd) mirror neutrinos in \cite{delAguila:2008zu}. 
	(The necessary Feynman rules and loop contributions are reviewed elsewhere.)
	
	Gauge invariance reduces the $\ell \to \ell' \gamma$ vertex for an on-shell photon 
	to a dipole transition,
	\bea
	\label{vertexmue} 
	i\ \Gamma^\mu_\gamma (p_\ell, p_{\ell'}) = i\ e \left[ i F_M^\gamma (Q^2) + F_E^\gamma (Q^2) \gamma_5\right] 
	\sigma^{\mu \nu}\ Q_\nu\ , 
	\eea
	where $Q_\nu = (p_{\ell'} - p_\ell)_\nu$. 
	Being the decay width (neglecting $m_{\ell'} (\ll m_{\ell})$) 
	\bea
	\label{decaywidthmue} 
	\Gamma (\ell \to \ell' \gamma) = \frac{\alpha}{2}\ m_\ell^3\ ( |F_M^\gamma|^2 + |F_E^\gamma|^2 )\ ,  
	\eea
	where $\alpha = e^2 /4 \pi$, 
	and the form factor (defining $\alpha_W = \alpha / s_W^2$)  
	\begin{align}
		\label{dipoleformfactormue} 
		F_M^\gamma  = 
		\theta_{\ell' j} \theta^*_{\ell j}\ \frac{\alpha_W}{16\pi}\frac{m_{\ell}}{M_{W}^2}  F^\chi_{M}\left(\frac{M_{W}^2}{M_j^2}\right)\ ,
	\end{align}
	with 
	\begin{align}
		\label{dipoleformfactormue1} 
		F^\chi_M (x) & = - \frac{2 + 5x - x^2}{4(1-x)^3}-\frac{3x}{2(1-x)^4}\ln x 
		\quad \stackrel{{x\rightarrow 0}}{\longrightarrow} \quad -\frac{1}{2}\  , 
	\end{align}
	and $F_M^\gamma = -i F_E^\gamma$. 
	While the corresponding branching ratio reads 
	\footnote{A lot of attention has been payed to this process in the past 
		\cite{delAguila:1982yu,Hisano:1995cp,Illana:2002tg,Arganda:2005ji} 
		due to the stringent experimental bound on $\mu \to e \gamma$. 
		The contribution of the heavy (quasi--Dirac) neutrinos involves 
		the couplings in Eqs. (\ref{SMcurrents}) and (\ref{lhchargedcurrents}) as well as  
		the couplings accounting for the Goldstone boson exchange. This has been 
		calculated quite a few times in the past \cite{Bjorken:1977br,Cheng:1977nv,Lim:1981kv,Vergados:1985pq,GonzalezGarcia:1991be,Ilakovac:1994kj,Tommasini:1995ii,Illana:2000ic}, together with other LFV transitions.} 
	\begin{align}
		\label{branchingratio} 
		{\rm Br} (\ell \to \ell' \gamma) = \frac{3\alpha}{2\pi} \left|\,\theta_{\ell' j} \theta^*_{\ell j} F^\chi_M\left(\frac{M_{W}^2}{M_j^2}\right)\right|^2 \ . 
	\end{align}
	Then, in order to estimate the bounds on the mixing we can substitute $F_M^\chi$ by 
	its limit $-1/2$ for $M_j^2 \gg M_W^2$ (see Eq. (\ref{dipoleformfactormue1})), 
	resulting in the bounds in Table \ref{Limits}. 
	If we further assume the moduli of $\theta_{\mu k}$ and $\theta_{e k}$ to be less than 0.03 
	(see Table \ref{Limits}), 
	as indicated by EWPD, they must be aligned with a precision higher than 2.4 \% to fulfill 
	the LFV bound on $\mu \to e \gamma$. 
	No similar (significant) constraint can be derived from $\tau$ decays at present. 
	
	Although it is flavor conserving, we can also compute the contribution to the 
	muon magnetic moment $a_\mu = 2 m_\mu F_M^\gamma$ (see \cite{delAguila:2019htj} for the 
	contribution of T--odd leptons), 
	whose current experimental value is $a_\mu^{\rm  exp} = (116592091\pm63)\times 10^{-11}$
	~\cite{Tanabashi:2018oca}. 
	With the same assumptions as above 
	$\delta a^{\rm T-even}_\mu = - 1.2 \times 10^{-9}\ \theta_{\mu k} \theta_{\mu k}^*$ 
	and then, equal to $- 1.1 \times 10^{-12}$ for $\theta_{\mu k} \theta_{\mu k}^* = (0.03){\tiny }^2$. 
	Which is too small (and negative) to explain a significative departure from the SM prediction, 
	$a_\mu^{\rm  SM} = (116591823\pm43)\times 10^{-11}$ 
	\cite{Tanabashi:2018oca}. 
	
	Similarly to the T--odd contribution to the EDM$_e$, $d_e = -e F^\gamma_E$, 
	the contribution of the heavy quasi-Dirac neutrinos vanishes at one loop. 
	A full two--loop calculation \cite{Barr:1990vd} is beyond the scope of this paper, 
	although its current experimental precision $|d_e| < 1.1 \times 10^{-29}$ $e$--cm at 90 \% C.L. 
	\cite{Andreev:2018ayy} merits it. 
	
	\section{Conclusions}
	\label{Conclusions}
	
	The LHT is a phenomenologically viable model with a composite Higgs. 
	It is minimal in the sense that all other (pseudo--) NG bosons are T--odd, 
	as there are the extra gauge bosons and almost all extra fermions, 
	while all SM fields are T--even. 
	This translates into less stringent constraints on their indirect effects 
	and on their direct production because they have to be always pair--produced. 
	
	Our long--term goal is to automate the calculation of the phenomenological 
	predictions of a definite LHT model which can be 
	confronted to experiment, as the minimal supersymmetric scenarios, and 
	guide collider searches. This means fixing the minimal fermion content 
	that makes the experimentally most restrictive processes one--loop finite 
	while keeping the Higgs boson mass free from quadratic divergences. 
	This concerns the quark as well as the lepton sector, and in this latter 
	case the charged LFV processes which are the most stringently constrained. 
	The contributions of the T--odd (heavy) leptons in the standard construction 
	which are necessary to make the Higgs decays finite are calculated in Refs. 
	\cite{delAguila:2017ugt,delAguila:2019htj}. In order to make the Higgs boson 
	mass free of quadratic divergences one must also include the SM singlets 
	in the RH $SO(5)$ quintuplets. The contributions of these heavy quasi--Dirac 
	neutrinos to charged LFV transitions are reviewed elsewhere. 
	In this paper we point out that such a working model can also accommodate 
	neutrino masses and mixings as these heavy neutrinos allow to implement 
	the inverse see--saw mechanism in a natural way. 
	
	If they are chosen to be T--even, they mix at tree level with the SM neutrinos, 
	giving rise to a rich phenomenology which has attracted a lot of attention in the past  \cite{Abada:2007ux,Arganda:2014dta,DeRomeri:2016gum,Ballett:2019bgd,delAguila:2007qnc,delAguila:2008hw,Das:2014jxa,Dib:2017iva,Das:2017zjc,Ruiz:2017yyf,Cai:2017mow,Pascoli:2018heg}. 
	In the LHT, however, the parameters describing this mixing are common 
	to other sectors of the theory and this is then 
	further constrained by the corresponding experimental observables, in particular, 
	by EWPD and LHC production limits 
	\cite{Reuter:2013iya,Dercks:2018hgz}. 
	This inverse see--saw mechanism of type I does not need to break 
	T--parity, in contrast with the see--saw of type II induced by a non--vanishing 
	vev of the neutral component $\langle \Phi^0 \rangle$ of the pseudo--NG scalar triplet of hypecharge $-1$ present in the model. 
	Moreover, the induced contribution in this latter case is higher order 
	in the LHT Weinberg operator expansion and hence, further suppressed. 
	In any case LN must be explicitly broken. A breaking which we 
	assume to be small and deferred to the heavy LH $SU(5)$ singlet counterpart, $\chi_L$, of the 
	SM RH singlets, $\chi_R$, that live in $SO(5)$ quintuplets. (If alternatively $\chi_R$ is chosen to be T--odd, 
	the minimal coupling giving masses to the SM charged leptons has to be generalized to maintain the 
	LFV Higgs decay into fermion pairs one--loop finite.)  
	
\begin{figure}
		\centering
		\includegraphics[scale=0.3]{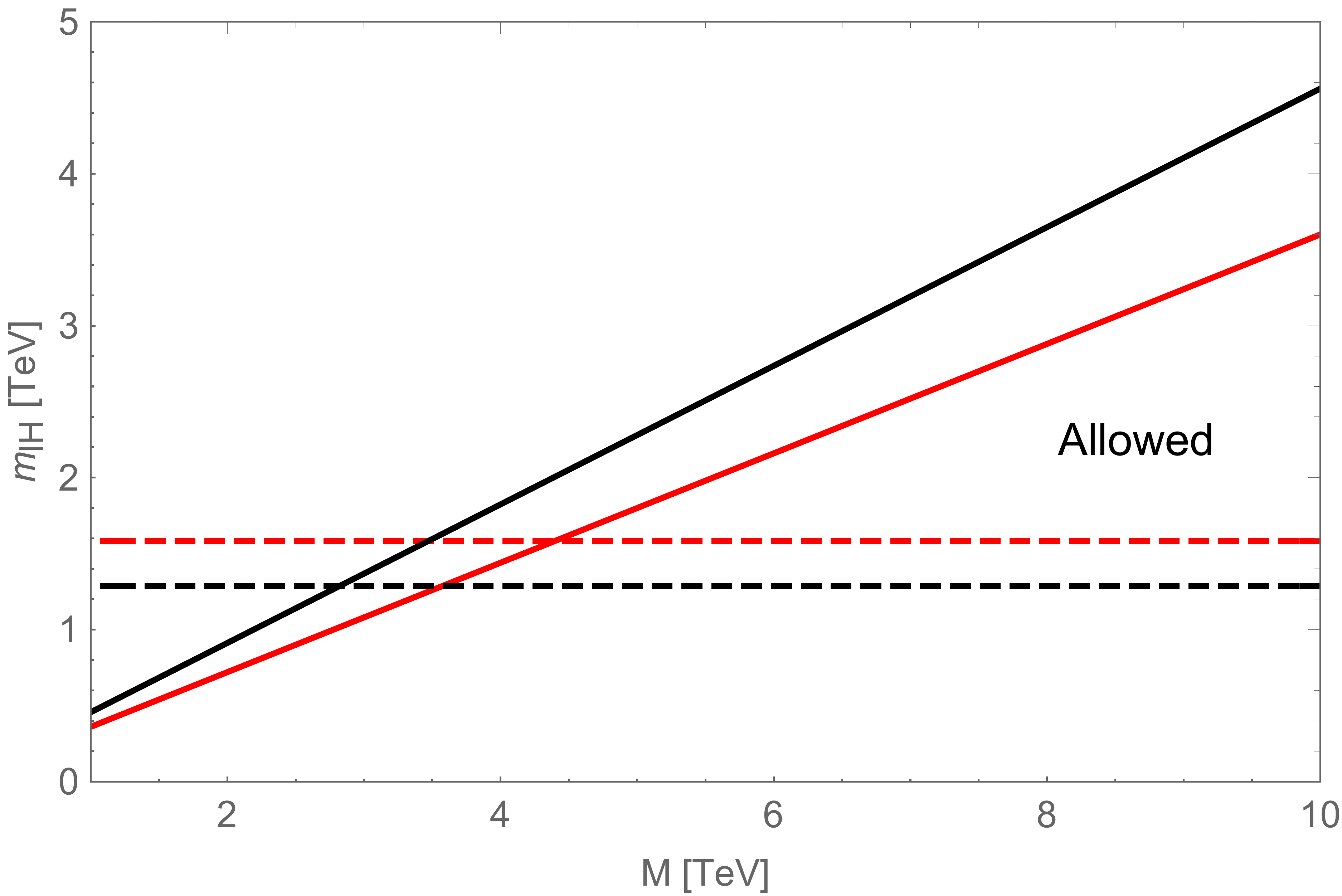} 
		\caption{Allowed mass region for the mirror lepton mass $m_{\ell H}$ versus the 
			quasi--Dirac neutrino mass $M$ for different values of the next physics scale $f$. Solid 
			lines are fixed by the upper bound of 0.03 on the mixing between SM and heavy neutrinos 
			for $f=1.5$ TeV (red) and 1.9 TeV (black). While dashed lines delimit the regions excluded by 
			the non--observation of mirror leptons.}
		\label{plot}
	\end{figure}

	As already emphasized, current experimental limits on the allowed departure from the SM predictions can
	be easily accommodated by the relatively large number of parameters fixing the LHT.
	Further fine tuning in the neutrino sector is only necessary for the LFV mixing, 
	which has to be typically adjusted to $1\%$ for $\mu$ to $e$ transitions. 
	Nevertheless, the expected range of variation of the LHT parameters 
	makes quite interesting future searches at the LHC.
	In the inverse see--saw mechanism the observed neutrino
	masses and mixings are uncorrelated, in the absence of a flavor symmetry, 
	with the masses of the heavy quasi--Dirac neutrinos and their
        mixing with the light sector.~\footnote{Flavor symmetries
          based on
          $A_4$ or $S_3$ (see for
          instance~\cite{
           Csaki:2008qq,Hirsch:2009mx,delAguila:2010vg,Altarelli:2010gt,Ma:2014qra}) 
          could be implemented to predict the observed pattern of
          lepton masses and mixing angles preventing at the same time
          large LFV transitions. However, this goes beyond the scope of this
          article and it is postponed to future work.}  
	In fact, the small LN violating masses $\mu$ for $\chi_L$ in Eq. (\ref{Majorana}) 
	can be adjusted to reproduce the light neutrino masses and mixings for 
	any (non--singular) value of the heavy--light mixing 
	(see Eq. (\ref{dlightmasses})). 
	Quasi--Dirac neutrino masses and mixings are only bounded on the other hand 
	by their direct production limit, which for $M$ is currently of the order of $M_W$ 
	\cite{Sirunyan:2018mtv},
	\footnote{Quasi--Dirac neutrinos are mainly produced by the exchange of 
		$W^{\pm}$, $Z$ and $h$ at the LHC (see Eqs. (\ref{lhchargedcurrents}) and (\ref{lhcurrents})) 
		\cite{Ruiz:2017yyf}. 
		But these amplitudes are proportional to the heavy neutrino mixing with the SM neutrinos 
		and it must fixed to its current upper bound of 0.03 to maximize the 
		direct production lower bound on $M$. 
		Besides, LN is practically conserved and the corresponding backgrounds are larger 
		than in the case of heavy Majorana neutrinos. 
		The most significant final states turn out to be three charged leptons plus missing energy 
		\cite{delAguila:2007qnc,delAguila:2008hw,Das:2014jxa,Dib:2017iva}, 
		and the expected reach for $M$ of the order of 300 GeV at the HL--LHC 
		\cite{Pascoli:2018heg}.}
	and by the non--observation of any significant departure from the SM 
	predictions in the leptonic sector. 
	The common dependence on the Yukawa coupling $\kappa$ in Eq. (\ref{mirror}) 
	of the mixing $\theta \simeq (v/\sqrt{2}) \kappa M^{-1}$ between the light and heavy 
	neutrinos and of the (T--odd) mirror lepton masses $m_{\ell H} \simeq \sqrt{2} \kappa f$ 
	delimits the $M-m_{\ell H}$ region allowed by the bound on $\theta < 0.03$. 
	Region, which is further restricted by the non-observation of heavy lepton 
	production{\tiny } \cite{Dercks:2018hgz}. 
	In Fig. \ref{plot} we draw these regions for $f = 1.5$ and $1.9$ TeV, red and black 
	lines, respectively. 
	In both cases quasi--Dirac neutrino masses below few TeV are excluded. 
	It must be emphasized, however, that the $m_{\ell H}$ production 
	limit depends on $f$ dramatically because pair production of 
	new vector--like leptons decaying into a SM lepton and the lightest T--odd boson (missing energy) at the LHC 
	is very much suppressed for $f > 2$ TeV \cite{Dercks:2018hgz}, 
	then drastically relaxing the lower bound on $M$. 
	The limit from neutrino mixing will improve with a more precise determination 
	of the constraints from EWPD while the improvement of the bound on lepton pair--production will 
	mainly require a higher colliding energy. 
	Both will cut down the allowed mass region in the LHT as a function of the 
	new physics scale $f$, mainly fixed by the non--observation of new (T--odd) gauge bosons. 
	More stringent limits on $f$ can be also derived from mirror quark production but as a function of their own Yukawa couplings \cite{Dercks:2018hgz}. 
	
	\section*{Acknowledgments}
	
	We thank previous work with Ll. Ametller, P. Talavera and R. Vega-Morales 
	and ongoing collaboration with T. Hahn, 
	and useful comments by J. Hubisz and I. Low. 
	This work has been supported in part by the Ministry of Science, Innovation and Universities, 
	under grant numbers FPA2016-78220-C3-1,2,3-P (fondos FEDER), 
	and by the Junta de Andaluc{\'\i}a grants FQM 101 and SOMM17/6104/UGR.

\end{document}